\documentclass[useAMS,usenatbib]{mn2e}
\usepackage{times,epsfig}
\usepackage{graphicx}
\usepackage{amsmath}
\newcommand{\mpcoh}{\,h^{-1}\,{\rm Mpc}}
\newcommand{\bs}{\mathbf{s}}
\newcommand{\br}{\mathbf{r}}
\newcommand{\bx}{\mathbf{x}}
\def\lsim{\,\lower2truept\hbox{${<\atop\hbox{\raise4truept\hbox{$\sim$}}}$}\,}
\def\gsim{\,\lower2truept\hbox{${> \atop\hbox{\raise4truept\hbox{$\sim$}}}$}\,}
\def\simlt{\mathrel{\rlap{\lower 3pt\hbox{$\sim$}}
        \raise 2.0pt\hbox{$<$}}}
\def\simgt{\mathrel{\rlap{\lower 3pt\hbox{$\sim$}}
        \raise 2.0pt\hbox{$>$}}}

\begin{document}

\title[Redshift-Space Distortions at Wide Angles]
{Simulating Redshift-Space Distortions for Galaxy Pairs with Wide Angular Separation}
\author[A. Raccanelli et al.]{
\parbox[t]{\textwidth}{
  Alvise Raccanelli$^{1}$\thanks{e-mail: alvise.raccanelli@port.ac.uk (AR)},
  Lado Samushia$^{1,2}$, 
  Will J. Percival$^{1}$}
\vspace*{4pt} \\
$^{1}$Institute of Cosmology \& Gravitation, University of Portsmouth,
Portsmouth, PO1 3FX, UK \\
$^{2}$National Abastumani Astrophysical Observatory, Ilia State University, 2A Kazbegi Ave, GE-0160 Tbilisi, Georgia}

\date{\today} 
\pagerange{\pageref{firstpage}--\pageref{lastpage}} \pubyear{2010}
\maketitle
\label{firstpage}

\begin{abstract}

  The analysis of Redshift-Space Distortions (RSD) within galaxy
  surveys provides constraints on the amplitude of peculiar velocities
  induced by structure growth, thereby allowing tests of General
  Relativity on extremely large scales. The next generation of galaxy
  redshift surveys, such as the Baryon Oscillation Spectroscopic
  Survey (BOSS), and the Euclid experiment will survey galaxies out to
  $z=2$, over $10,000$--$20,000\,{\rm deg}^2$. In such surveys, galaxy
  pairs with large comoving separation will preferentially have a wide
  angular separation. In standard plane-parallel theory the
  displacements of galaxy positions due to RSD are assumed to be
  parallel for all galaxies, but this assumption will break down for
  wide-angle pairs.  \citet{szapudi04} and \citet{papai08} provided a
  methodology, based on tripolar spherical harmonics expansion, for
  computing the redshift-space correlation function for all angular
  galaxy pair separations. In this paper we introduce a new procedure
  for analysing wide-angle effects in numerical simulations. We are
  able to separate, demonstrate, and fit each of the effects described
  by the wide-angle RSD theory. Our analysis highlights some of the
  nuances of dealing with wide-angle pairs, and shows that the effects
  are not negligible even for relatively small angles. This analysis
  will help to ensure the full exploitation of future surveys for RSD
  measurements, which are currently confined to pair separations less
  than $\sim80\mpcoh$ out to $z\simeq0.5$.

\end{abstract}


\section{Introduction}

Many different mechanisms have been suggested to explain the observed
late-time acceleration of the expansion of the Universe
\citep{riess98,perlmutter99}.\footnote{For a recent review of dark
  energy see, e.g., \citep{frieman08} and references therein.}
Differentiating between these options is one of the main challenges
facing cosmologists today. We can try to build up the evidence for
different mechanisms by examining the evolution of the Universe in two
key ways: measuring the background geometry and measuring structure
formation within it.

The geometrical evolution of the Universe can most easily be measured
using two primary techniques: we can use supernovae as standard
candles or galaxy clustering as a standard ruler to make precise
measurement of cosmological expansion. Although supernovae were first
to confirm the accelerated expansion of the Universe at high
statistical significance \citep{riess98,perlmutter99} following
analyses of early galaxy surveys \citep{efstathiou90}, supernovae
surveys are now limited by systematic rather than statistical errors
\citep{kessler09}. The possibility of using galaxy clustering to
provide a standard ruler has become increasingly important since the
baryon acoustic peak was detected
\citep{percival01,cole05,eisenstein05} in galaxy power spectra
measured from the 2dF Galaxy Redshift Survey (2dFGRS;
\citealt{colless03}) and the Sloan Digital Sky Survey (SDSS;
\citealt{york00}). Using only the Baryon Acoustic Oscillation (BAO)
component of the galaxy clustering signal makes constraints robust to
non-linear effects, and has already been exploited to produce
interesting constraints on cosmological models
\citep{percival07a,percival07b,gaztanaga08,sanchez09,percival10}.

Galaxy surveys provide complementary information about the build-up of
large-scale structure through Redshift Space Distortions (RSD). These
arise because we do not observe true galaxy positions, but instead
infer distances from measured redshifts. Coherent comoving galaxy
velocities due to the growth of structure therefore lead to measurable
anisotropic clustering \citep{kaiser87,hamilton98}. RSD have been
measured from the 2dFGRS and SDSS using techniques based on both
correlation functions and power-spectra \citep{peacock01, hawkins03,
  percival04, pope04, zehavi05, okumura08, cabre09}, and have recently
been detected at higher redshift \citep{guzzo09,blake10}.

In fact, for fundamental reasons, RSD are independent of galaxy bias:
galaxies act as test particles in the matter flow, so their motion is
independent of galaxy properties. We can therefore measure the matter
velocity field at the locations of the galaxies and, provided that
galaxy positions are representative within the velocity field, this
gives an unbiased measurement of $f\sigma_8({\rm mass})$ where
$f\equiv d\ln D/d\ln a$, the logarithmic derivative of the linear
growth rate, $D(z)$, with respect to the scale factor $a$, and
$\sigma_8({\rm mass})$ quantifies the amplitude of fluctuations in the
matter density field.

Many techniques have been proposed for measuring RSD, including using
multipoles of the correlation function \citep{hamilton92}, or the
power spectrum \citep{percival09}, based on the plane-parallel
approximation for RSD \citep{kaiser87}. The geometry of the system can
be easily incorporated into an analysis of the correlation function as
the separation of each pair can be split into radial and angular
components. It is more difficult to perform such a decomposition when
measuring the power spectrum, although various techniques have been
suggested, decomposing the density field into a basis of spherical
harmonics and radial functions (e.g. \citealt{fisher94,heavens95}). As
well as allowing for the geometry to a survey and RSD within it, we
need to allow for two effects:
\begin{enumerate}
\item RSD are degenerate with the angular anisotropy of power spectrum
  caused by the Alcock-Palczinski effect, so measurements of growth and
  geometry from the same survey will be correlated, and it is sensible
  to undertake a combined analysis
  \citep{ballinger96,simpson09}. 
\item as noted by \citet{hamilton98}, the full RSD operator has extra
  terms compared to the plane-parallel one and results in a
  redshift-space power spectrum which is not diagonal in
  ${\bf k}$. These extra terms are important for galaxy pairs
  separated by wide angles and need to be included in any analysis
  that uses wide-angle data \citep{szalay98,szapudi04,papai08}.
\end{enumerate}
In this paper we concentrate on the second of these effects.
\citet{szapudi04} and \citet{papai08} proposed a method of computing
the redshift-space correlation function which does not assume a
plane-parallel approximation and is applicable to arbitrarily large
angles. In their treatment the redshift-space correlation function
depends not only on the redshift-space pair separation $x$ and cosine
of the angle of the pair with respect to the line-of-sight $\mu$, but
also on the separation angle between two galaxies in a pair $\theta$,
which does not have to be small. They were able to express the
redshift-space correlation function explicitly trough a real-space
correlation function and cosmological parameters.  Given the precision
that future surveys will achieve and the fact that they will cover large
fractions of the sky, these effects will need to be corrected.

The wide-angle effects can be subdivided in ``purely wide-angle'' and
``mode-coupling'' terms: ``purely wide-angle'' effects correct
plane-parallel predictions accounting for the fact that the separation
angle is non-zero, ``mode-coupling'' terms in addition account for the
fact that galaxy pairs coherently moved from the high-density to
low-density regions. These latter terms vanish if the initial
real-space distribution of galaxies is uniform in distance, and both
terms vanish in the plane-parallel limit due to the symmetry of the
system.  Both terms are of the same order and mode-coupling tends to
smooth out features in the power spectrum such as the Baryon Acoustic
Oscillations (BAO).

Given the importance and, particularly, the quality of RSD constraints
expected from forthcoming surveys such as the Baryon Oscillation
Spectroscopic Survey (BOSS; \citealt{schlegel09a}), surveys resulting
from proposals for wide-field multi-object spectrographs on 4m
telescopes such as BigBOSS \citep{schlegel09b} and satellite missions
such as Euclid \citep{laureijs09}, it is timely to revisit this
problem, and to test wide-angle RSD theory. In this paper we use
numerical simulations to extensively test the dependence of RSD
constraints on the angular separation of galaxy pairs.

This paper is organised as follows: in Sec.~\ref{sec:rsd} we briefly
review the theory of wide-angle RSD; in Sec.~\ref{sec:rsdmeas} we
describe a time-efficient method of getting a low-noise measurement of
wide-angle correlation function from a mock HV catalog; in
Sec.~\ref{sec:results} we present our results and discuss all the
steps necessary to match measured correlation function with
theoretical predictions; in Sec.~\ref{sec:disc} we conclude and
discuss how the results will affect real survey measurements.

\section{Wide-Angle Redshift-Space Distortions}
\label{sec:rsd}

Galaxy positions are measured in a redshift-space, and differ from the
real-space positions because of the contributions from peculiar
velocities,
\begin{equation}
  \bs(\br) = \br - v_r(\br) \hat{\br},
\end{equation}
where $\bs$ is the redshift space position, $\br$ is the real space
position and $v_r$ the radial component of peculiar velocity.

Since the total number of galaxies in real and redshift spaces is the
same, the number $N^s(\bs)d^3s$ of galaxies observed in a volume element
$d^3s$ of redshift space is related to the real space number density
$N^r(\br)$ by:
\begin{equation} \label{eq:nd}
  N^s(\bs) d^3 s = N^r(\br) d^3 r.
\end{equation}
The observed redshift space galaxy overdensity $\delta^s$ at position
$\bs$ can be related to the redshift-space selection function
$\bar{N}^s$,
\begin{equation}
  \delta^s(\bs) = \frac{N^s(\bs) - \bar{N}^s(\bs)}{\bar{N}^s(\bs)}.
\end{equation}
In contrast, an unbiased estimate of the true galaxy overdensity
$\delta(\br)$ at position $\br$ is given by:
\begin{equation}
  \delta^r(\br) = \frac{N^r(\br) - \bar{N}^r(\br)}{\bar{N}^r(\br)}, 
\end{equation}
where $\bar{N}^r(\br)$ is the expected galaxy distribution in
real-space -- the expected number density of unclustered galaxies at
position $\br$ given the selection criteria of the survey.

We distinguish here between the measured redshift-space selection
function $\bar{N}^s(\bs)$ and the true selection function
$\bar{N}^r(\br)$ even though the two are the same at linear order. We
now assume that the expected density is a function of the radial
component of the position only, which we denote $s$ in redshift-space
and $r$ in real-space. Thus the relation between the observed redshift
space overdensity $\delta^s(\bs)$ and the true overdensity
$\delta^r(\br)$ is:
\begin{equation}
  \bar{N}^s(s) [1+\delta^s(\bs)] s^2 ds = \bar{N}^r(r) [1+\delta^r(\br)] r^2 dr.
\end{equation}
Using Eq.~(\ref{eq:nd}) we obtain:
\begin{align}
  1+\delta^s(\bs) = & [1+\delta^r(\br)] 
    \left(1+ \frac{\partial v}{\partial r} \right)^{-1} \\ \nonumber
    & \left(1+ \frac{v}{r} \right)^{-2} \frac{\bar{N}(r)}{\bar{N}(r+v\hat{r})},
\end{align}
which gives, to linear order:
\begin{equation}
  \delta^s(\br) = \delta^r(\br)-
    \left( \frac{\partial v}{\partial r}+\frac{\alpha(\br)v}{r} \right),
\end{equation}
where 
\begin{equation}
  \alpha(\br) = \frac{\partial \ln r^2 \bar{N}^s(\br)}{\partial \ln r}.
\end{equation}
Usually the $(1+\frac{v}{r})^{2}$ term in the Jacobian is omitted
because, in the linear regime, it gives rise to a $2v/r$ term, that
would tend to zero at large distances. However, \citet{papai08} have
argued that, for wide angles, the $v/r$ term is of the same order as
the $\partial_r v$ term. 

\begin{figure}
\centering
\includegraphics[width=0.8\columnwidth]{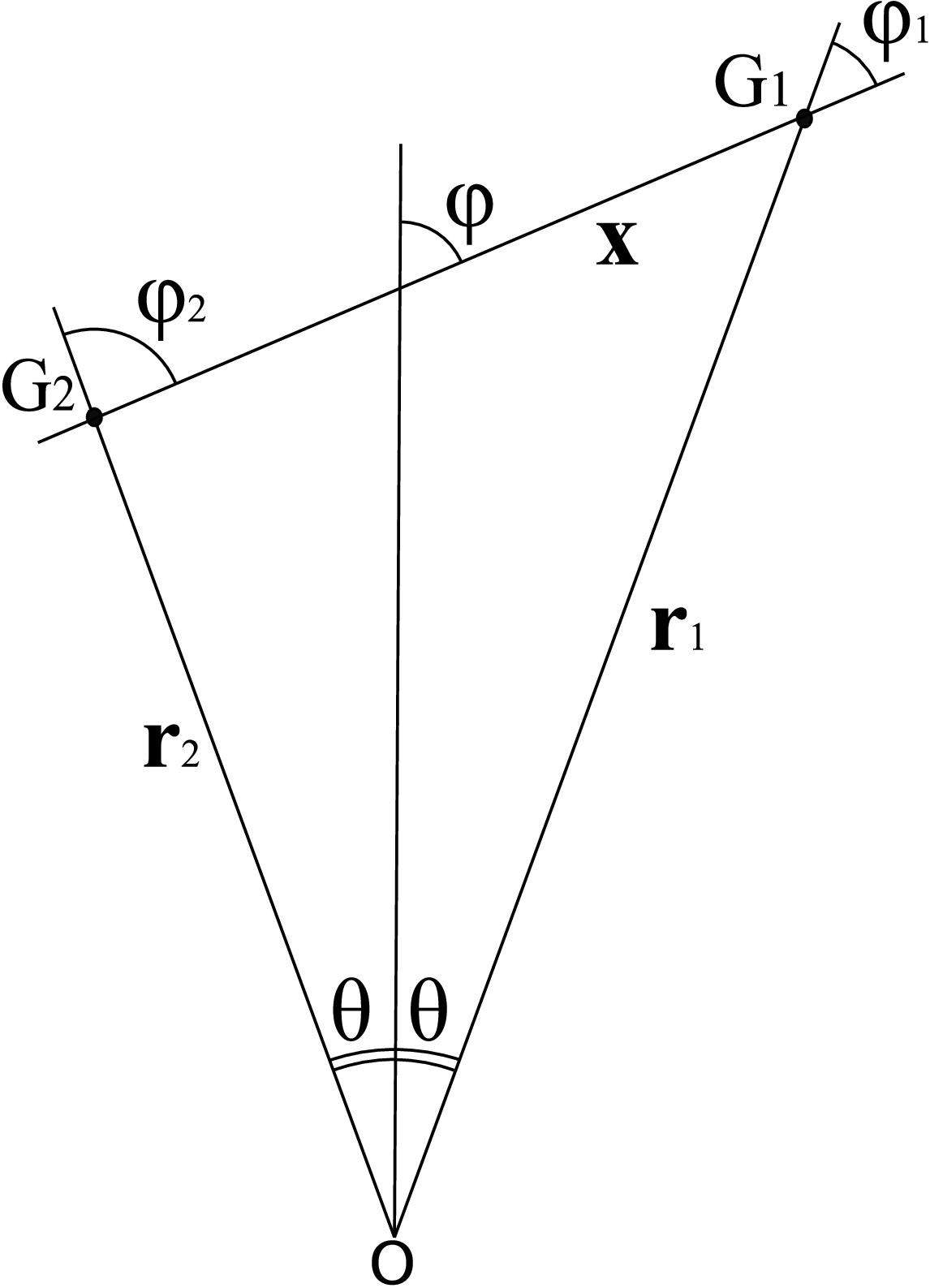}
\caption{The coordinate system adopted for the triangle formed by the
  observer $O$, and galaxies $G_1$ and $G_2$. \label{fig:triangle}}
\end{figure}

Using the exact Jacobian, and following \citet{papai08}, we can express
the linear overdensity as:
\begin{equation}  \label{eq:del}
  \delta^s(r) = \int \frac{d^3k}{(2\pi)^3} e^{i k_j r_j}
    \left[1+f(\hat{r}_j\hat{k}_j)^2 
      - i\alpha f\frac{\hat{r}_j\hat{k}_j}{rk} \right]
    \delta(k),
\end{equation}
so the correlation function is:
\begin{align}  
  \langle \delta^s(\mathbf{r}_1) \delta^{s*}(\mathbf{r}_2) \rangle &= \int \frac{d^3k}{(2\pi)^3} P(k) e^{ik(r_1-r_2)} \nonumber \\ 
  & \left[ 1 + \frac{f}{3} + \frac{2f}{3} L_2(\hat{r}_1 \hat{k}) 
    - \frac{i\alpha f}{r_1 k} L_1(\hat{r}_1 \hat{k}) \right] \nonumber \\
  & \left[ 1 + \frac{f}{3} + \frac{2f}{3} L_2(\hat{r}_2 \hat{k}) 
    + \frac{i\alpha f}{r_2 k} L_1(\hat{r}_2 \hat{k}) \right] , \label{eq:xi1}
\end{align}
where $L_{\ell}$ are the Legendre polynomials of order $\ell$ and $P(k)$ is
the real-space matter power spectrum. The third terms in the brackets
are the ones responsible for the wide-angle effects, while the fourth terms are
the ones responsible for the mode-coupling. Note that the $r_1$ and $r_2$ 
terms in the denominator depend on the angular separation of the galaxies.  
Setting $\phi_1$ as the angle between the vector to the first galaxy in a pair 
$\bf{r}_1$, $\bf{x}$ to be the vector connecting galaxies in a pair, and $\phi_2$ 
to be the angle between vector to the second galaxy in a pair $\bf{r}_2$ and
$\bf{x}$, we can use the sine rule (see Fig.~\ref{fig:triangle}) to
express $r_1$ and $r_2$ as
\begin{align}
  &r_1 = \frac{\sin(\phi_2)}{\sin(\phi_2-\phi_1)} x, \label{eq:g1} \\
  &r_2 = \frac{\sin(\phi_1)}{\sin(\phi_2-\phi_1)} x. \label{eq:g2}
\end{align}

Tripolar spherical harmonics are the most natural basis for the expansion of a function that depends on three directions
\citep{varshalovich88} so, as suggested in \citet{szapudi04} and \citet{papai08}, we expand Eq.~(\ref{eq:xi1})
using a subset of them that are proportional to the zero angular momentum:
\begin{align}
&\ S_{\ell_1 \ell_2 \ell}(\hat{r}_1, \hat{r}_2, \hat{x}) = \\
&\ = \sum_{m_1,m_2,m} \left( \begin{array}{ccc} \ell_1 & \ell_2 & \ell \\ \nonumber
m_1 & m_2 & m  \end{array} \right) C_{\ell_1m_1}(\hat{r}_1) C_{\ell_2m_2}(\hat{r}_2) C_{\ell m}(\hat{x}), \label{eq:S_exp}
\end{align}
where $C_{\ell m}$ are normalized spherical functions, multiplied by the $3j$ Wigner symbol.
The redshift space correlation function is then written as:
\begin{equation}
\xi^s(\hat{r}_1,\hat{r}_2,\hat{x}) = \sum_{\ell_1,\ell_2,\ell} B^{\ell_1\ell_2\ell}(x,\phi_1, \phi_2)S_{\ell_1\ell_2\ell}(\hat{r}_1, \hat{r}_2, \hat{x}), \label{eq:BS_form}
\end{equation}
where $B^{\ell_1\ell_2\ell}(x,\phi_1, \phi_2)$ are a series of coefficients that depend on $f$, $g_i(\phi_i)$ and $\xi^r_{\ell}(x)$, and they
can be divided into wide-angle
components (given in \citealt{szalay98} \& \citealt{szapudi04}) that
do not depend on the third term in Eq.~(\ref{eq:del}), and
mode-coupling components (given in \citealt{papai08}); 
the plane-parallel approximation emerges as a limit when $\hat{r}_1=\hat{r}_2$.

As described in \citet{hamilton96}, for any pair of galaxies we can
fully analyse the problem within a plane formed by the two galaxies
and the observer; the effect of redshift-space distortions only
depends on the geometry within this plane. In the coordinate system
described above, the vectors connecting the observer with the galaxies
are $\bf{r}_1$ and $\bf{r}_2$, and the separation vector between the
galaxies is $\bf{x}$; $\phi_i$ are the angles between $\bf{r}_i$ and
$\bf{x}$. In the plane parallel approximation $\phi_1=\phi_2$. This
description of the triangle is shown in Fig.~\ref{fig:triangle}.

At this point we still have the freedom to choose a line-of-sight to
the galaxy {\em pair}: we follow the standard approach and choose the
line of sight as a direction bisecting the angle formed by the two
galaxies, which we assume is given by $2\theta$. The angle between the
angle bisector and $\bf{x}$ is denoted $\phi$, and we define
$\mu\equiv\cos(\phi)$. We therefore have that $\phi =
\frac{1}{2}(\phi_1+\phi_2)$, and $\theta =
\frac{1}{2}(\phi_2-\phi_1)$. This is shown in Fig.~\ref{fig:triangle}.

With this choice of coordinate system the redshift-space correlation
function can be expanded as:
\begin{eqnarray}
\label{eq:xiwa}
  \xi^s(x,\theta,\phi) &=& \sum_{ab} c_{ab}(x) 
    L_a(\cos(\theta)) L_b(\cos(\phi)),\\
  &=& \sum_{n_1,n_2} a_{n_1n_2} \cos(n_1\phi_1)\cos(n_2\phi_2) \nonumber\\
  && \hspace{1.2cm} + b_{n_1n_2} \sin(n_1\phi_1)\sin(n_2\phi_2),
\end{eqnarray}
with coefficients given in \citet{papai08}. Eq.~(\ref{eq:xiwa})
reduces to the plane-parallel result \citep{hamilton92} in the limit
of $\theta\rightarrow0$.

\section{Measuring Wide-Angle Correlation Functions}
\label{sec:rsdmeas}

A standard method for analysing a N-body simulation to test wide-angle
redshift-space distortions would be to:
\begin{enumerate}
  \item locate an observer within the output,
  \item translate all galaxies from real into redshift-space based on this observer,
  \item sample from these galaxies based on desired radial distribution,
  \item split pairs into bins in $\phi$, $\theta$, $x$ and counts pairs,
  \item estimate the correlation function.
\end{enumerate}
This approach mimics that of an actual survey analysis, creating a
mock galaxy catalogue. However, if we want to measure the effect of
redshift-space distortions for galaxy pairs with a particular angular
separation, the method is not optimal as only a small fraction of the
galaxy pairs analysed will have this angular separation. In addition,
we have to perform the full procedure for every radial galaxy
distribution that we wish to analyse, and this distribution will limit
the pair-density that can be selected from the simulation. In order to
rapidly increase the signal-to-noise, we adopt a different procedure,
allowing the origin to move so that each galaxy pair can be analysed
as if it was observed with the required angular separation. This
procedure can be summarised as follows:
\begin{enumerate}
  \item decide on the value of $\theta$ for which we wish to analyse pairs,
  \item take each galaxy pair from the simulation with real-space
    separation $<R_{\rm max}\mpcoh$,
  \item for each pair randomly choose $\mu\in[-\cos(\theta),\cos(\theta)]$,
  \item choose the location of the origin giving this $\mu$ and $\theta$,
  \item move galaxies according to their expected redshift-space distortion,
  \item weight the pair by a function of $\mu,x$ to match desired distribution,
  \item split pairs into bins in $\phi$, $x$ and counts pairs,
  \item estimate the correlation function.
\end{enumerate}
The added complexity of including redshift-space distortions on a
pair-by-pair basis is outweighed by the ability to obtain more pairs
at the desired angular separation. Note that $\mu$ in step (iii) is
constrained and cannot have any value within $[-1, 1]$ because
$|\phi|>|\theta|$ results in geometrically impossible triangles (see,
Fig.~\ref{fig:triangle}). The distributions of triangles in $\mu$ and
$x$ for a galaxy distribution with a power law selection in $r$ can be
calculated analytically, as we now demonstrate.

\subsection{Real-space distribution of $\mu$, $x$}

The redshift-space correlation function in Eq.~(\ref{eq:xiwa}) depends
on the selection function $\alpha(\br) \equiv 2
+ \partial\ln[n(\br)]/\partial r$, where $n(\br)$ is the number
density of galaxies in real space and $\br$ is the position of the
galaxy from the observer as in Section~\ref{sec:rsd}.  If the galaxy
distribution is uniform ($\alpha(\br) = 2$, $n(\br)=n$, both independent of ${\br}$), then the probability of finding a galaxy in some region is
proportional to the volume of that region. Therefore if we randomly pick
a galaxy in the catalog, there will be on average $n x^2 dx d\mu$
galaxies in a small volume element $dx d\mu$ which is distance $x\pm
dx$ away from that galaxy within an angular slice $\mu \pm d\mu$. In
the plane-parallel approximation, where all galaxies are assumed to lie
along the same direction from the observer, the distribution of galaxy
pairs will scale as
\begin{equation}
  dN^{\rm pair}_{\rm pp}(x,\mu)\propto x^2 dx d\mu,
\label{eq:dN}
\end{equation}
where $dN^{\rm pair}_pp(x,\mu)$ is the number of pairs with separation
$x\pm dx$ and cosine of the angle with the line of sight in the
$\mu\pm d\mu$ interval.  If we pick galaxy pairs separated by a fixed
opening angle from the uniform spatial distribution of galaxies (step
(i) in the procedure described above), for a large enough angle the
distribution of pairs will not follow Eq.~(\ref{eq:dN}).  For galaxy
pairs with a fixed half-angle $\theta$ (see, Fig.~\ref{fig:triangle}),
the likelihood of finding one galaxy at position ${\bf r_1}$ and
another at ${\bf r_2}$ is $P(r_1)\propto r_1^2dr_1$, $P(r_2)\propto
r_2^2dr_2$. Since the two likelihoods are independent, the joint
probability of finding that pair of galaxies is:
\begin{equation}
  dN^{\rm pair}_\theta(r_1,r_2) \propto r_1^2r_2^2dr_1dr_2.
\label{eq:Nxx}
\end{equation}
Using Eqs.~(\ref{eq:g1}) and (\ref{eq:g2}) we can rewrite
Eq.~(\ref{eq:Nxx}) in terms of $\phi$ and $x$ as:
\begin{equation}
\label{eq:dNwa}
  dN^{\rm pair}_\theta(x,\mu) \propto x^5
    \sin(\phi+\theta)^2\sin(\phi-\theta)^2dx d\phi.
\end{equation}
After completing steps (ii) and (iii) in the previous subsection, we
will have a distribution of galaxy pairs that is uniform in $d\mu =
\sin(\phi)d\phi$ and scales as $x^2 dx$ with separation. This
distribution of galaxy pairs does not correspond to $\alpha = 2$. If
we want to compare our data with the correlation function computed
from Eq.~(\ref{eq:xiwa}) for $\alpha = 2$, we have to weight our
galaxy pairs (step (iv) in the previous subsection) by an additional
factor of $x^3 \sin(\phi-\theta)^2\sin(\phi+\theta)^2$ to get a
distribution given by Eq.~(\ref{eq:dNwa}).

This procedure can be applied to the case of arbitrary $\alpha$. If,
for example, the galaxy number density scales as a power-law $n({\bf
  r})=r^{-N}$, then $\alpha = 2 - N$, and the distribution of galaxy
pairs with a fixed opening half-angle drawn from this distribution
will be:
\begin{equation}  \label{eq:dNwaN}
  dN^{\rm pair}_\theta(x,\mu) \propto x^{2\alpha+1}
    \sin(\phi+\theta)^{\alpha}\sin(\phi-\theta)^{\alpha}dx d\phi.
\end{equation}
We therefore weight all pairs recovered from the simulation with a
weight given by:
\begin{equation}  \label{eq:weight}
  W(x,\mu) = \frac{x^{2\alpha-1}
    \sin(\phi_r+\theta)^{\alpha}\sin(\phi_r-\theta)^{\alpha}}
  {x_{s}^{2\alpha-1}\sin(\phi_s+\theta)^{\alpha}\sin(\phi_s-\theta)^{\alpha}},
\end{equation}
where $x_{s}$ is the separation of the galaxies in redshift-space,
$\phi_r$ is the real-space value of $\phi$ for the chosen galaxy pair
and $\phi_s$ is the redshift-space value. We divide the real-space
distribution by the redshift-space equivalent in order to normalise
the weights to give no net change in the expected pair distribution.

\subsection{Redshift-space distribution of $\mu$, $x$}
\label{sec:rsdist}

When transformed into redshift space, the real-space galaxy
distribution is ``washed out'' by the random component of galaxy
velocities, so we need to convolve with the random velocity
distribution. This would normally not be a problem as we would
estimate the galaxy distribution directly in redshift-space. However,
using the above procedure we have set the real-space distribution of
galaxies, so we need to take care when modelling the redshift-space
distribution. To correct for this effect, we first estimate the
velocity dispersion $\sigma_v$ from the catalog, assuming that the
random velocities are drawn from a Gaussian distribution. We then
convolve the initial distribution in real space with this
Gaussian. For example, for $n({\bf r}) \propto r^{-N}$, the probability
of finding a galaxy at distance $r$ away from the observer is $P(r)\propto
r^{2-N} = r^\alpha$. In the unclustered redshift space this will
transform into:
\begin{eqnarray}
\label{eq:ps}
  P(s) &\propto& \displaystyle\int_0^{\infty}r^\alpha 
    \frac{e^{-(s-r)^2/2\sigma_v^2}}{\sqrt{2\pi\sigma_v^2}} \nonumber \\
  &\propto& (\sigma_v)^{\alpha+1}\left[\frac{s}{\sqrt{2}}\Gamma
    \left(\frac{\alpha+2}{2}\right)\,  
    _1 F_1\left(\frac{1-\alpha}{2},\frac{3}{2};
      -\frac{s^2}{2\sigma_v^2}\right)\right.\nonumber\\
  &+&\left.\frac{\sigma_v}{2}\Gamma\left(\frac{\alpha+1}{2}\right)\,  
    _1 F_1\left(\frac{-\alpha}{2},\frac{1}{2};
      -\frac{s^2}{2\sigma_v^2}\right)\right],
\end{eqnarray}
where F is a hypergeometric function.

For a specific case of $\alpha=2$, Eq.~(\ref{eq:ps}) results in:
\begin{equation}  \label{eq:psa2}
  P(s) \propto (\sigma_v^2+s^2)\left(1+{\rm Erf}
    \left(\frac{\sigma_v}{\sqrt{2}s}\right)\right)
    +\sqrt{\frac{2}{\pi}}\sigma_v se^{-s^2/2\sigma_v^2},
\end{equation}
\noindent
where $\rm Erf(r)$ is an error function. The distribution of galaxy
pairs for $\alpha=2$ will be:
\begin{equation}
  N^{\rm pair}_{\alpha = 2} \propto P(s_1)P(s_2) x_s dx_s d\phi,
\end{equation}
\noindent
where $P(s_i)$ is given by Eq.~(\ref{eq:psa2}) and $s_1$ and $s_2$ can
be expressed in terms of $\bs$ and $\phi$ via Eqs.~(\ref{eq:g1}) and
(\ref{eq:g2}), after the substitution of $\bx$ with $\bx_s$.

\begin{figure}
 \centering
  \includegraphics[width=1\columnwidth]{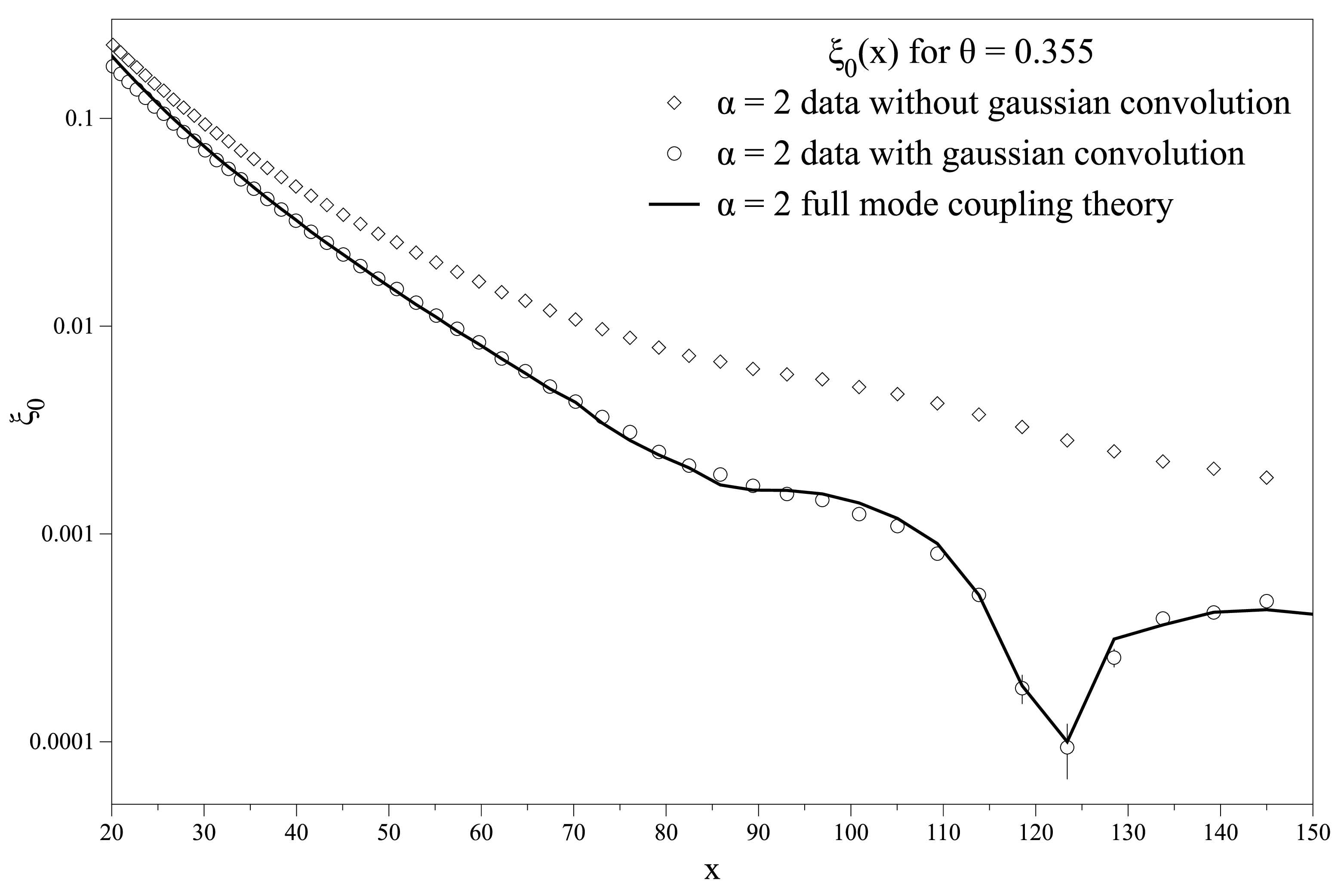}
  \caption{The monopole correlation function calculated from the HV
    mock catalogues designed to mimic a galaxy distribution with
    $\theta=0.355$, and $\alpha=2$. Open circles (with 1$\sigma$
    errors) were calculated using the redshift-space model
    for the galaxy distribution. Open diamonds used the real-space
    model. The solid line shows the prediction from
    Section~\ref{sec:rsd}. \label{fig:xi_cmpr_gauss} }
\end{figure}
When modelling the theoretical correlation function, \citet{hamilton96}
argue that the difference between $\bar{n}({\bf s})$ and $\bar{n}({\bf
  r})$ is small. While this is true when creating a model correlation function, it
is not true, and we need to use the $\bar{n}({\bf s})$ instead of
$\bar{n}({\bf r})$, when we model the data. This is shown in
Fig.~\ref{fig:xi_cmpr_gauss}, where we plot $\xi_0(x)$ calculated for
$\alpha=2$ from a mock catalog, assuming that the galaxy distribution
follows the one expected from either the real-space or redshift-space
calculation. See Section~\ref{sec:results} for more details of the
calculation.

\subsection{Pairs close to the origin}  \label{sec:close_pairs}

\begin{figure}
\centering
\includegraphics[width=0.9\columnwidth]{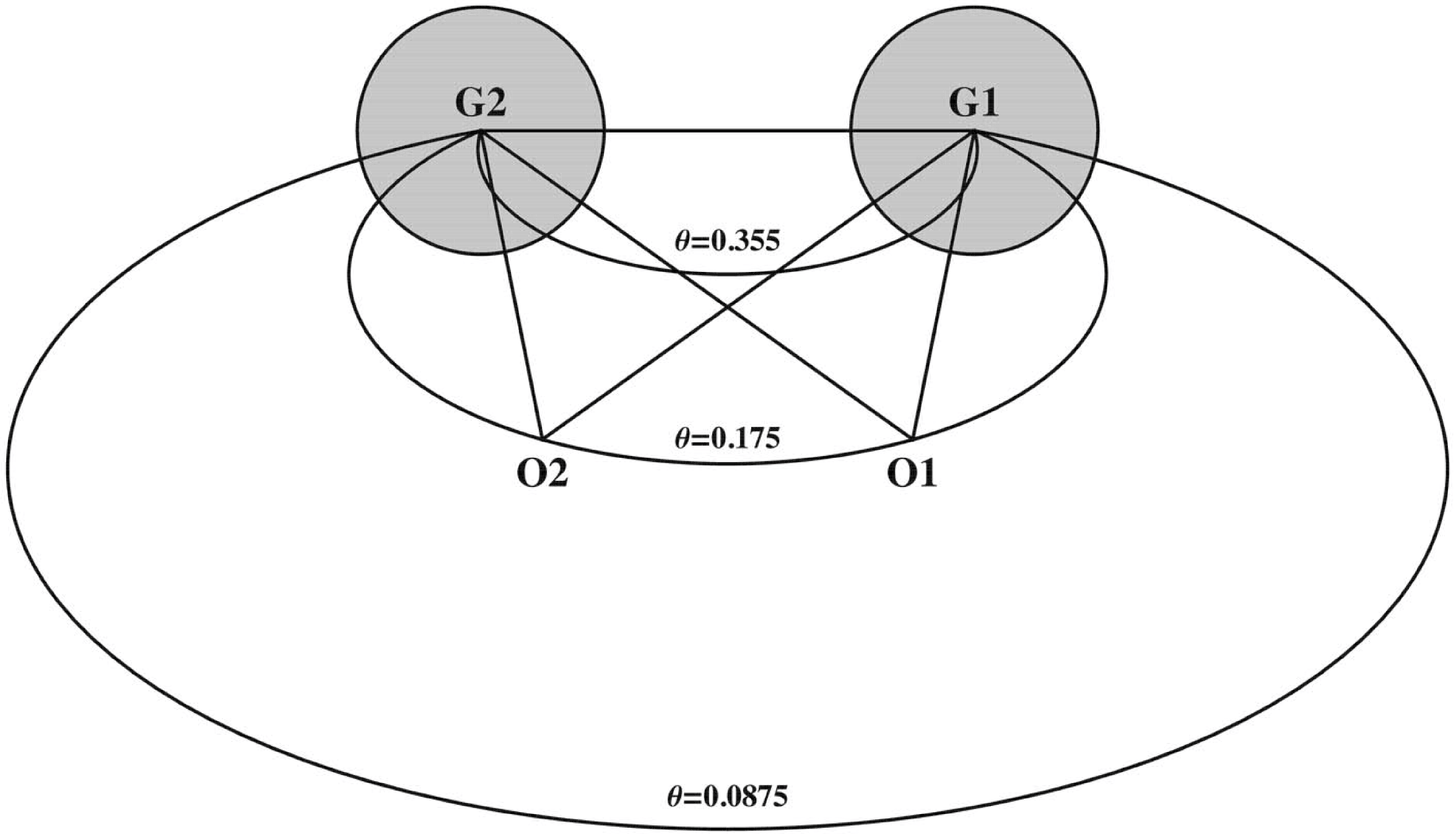}
\caption{Schematic representation showing the positions at which the
  origin could be placed for different values of $\theta$. The shaded
  circles around each galaxy are exclusion zones within which the
  redshift-space distortions is comparable with the distance to the
  galaxy. O1 and O2 mark two possible positions for an origin that
  would give angular galaxy separation of $2\theta=0.71$. \label{fig:excl}}
\end{figure}
Galaxies that are close to the origin can cause problems as, for such
galaxies, the redshift-space displacement can be larger than the
distance to the galaxy. This problem is exacerbated because we do not
include the velocity of the observer in our calculation. In extreme
situations, naively applying the expected redshift-space distortion
would place the galaxy on the opposite side of an observer. In order
to avoid such problems, we only include galaxy pairs where both
galaxies are more than $5\sigma_v$ away from the origin, where
$\sigma_v$ is the 1D velocity dispersion of the galaxy population. For
the HV simulation $\sigma_{v_{|z=0}}=3.9\mpcoh$. Fig.~\ref{fig:excl}
shows, for galaxies $G_1$ and $G_2$, the loci of positions at which
the origin could be placed for fixed $\theta$. The circles mark the
exclusion zones. If one of the galaxies in a pair is inside the
exclusion zone in real space, we do not include that pair when we
estimate the correlation function. This exclusion is tracked when we
calculate the expected galaxy distribution.

\subsection{The Hubble Volume Simulation}

We apply the procedure outlined in Sec.~\ref{sec:rsdmeas} to analyse
wide-angle redshift-space distortions within the $\Lambda$CDM Hubble
Volume (HV) simulation \citep{evrard02}. The $\Lambda$CDM HV
simulation, covering a $(3000\mpcoh)^3$ box, assumes a cosmological
model with $\Omega_m=0.3$, $\Omega_{CDM}=0.25$, $\Omega_b=0.05$,
$\Omega_{\Lambda}=0.7$, $h=70$, $\sigma_8=0.9$, \& $n_s=1$. We do not
apply any galaxy bias, simply Poisson sampling the matter particles to
give our ``galaxy distribution''; the inclusion of a bias model would
not alter the conclusions of this work. We use the periodic nature of
the numerical simulation to eliminate boundaries from our pair
counts. This means that, by using the above weighting scheme and
allowing for the removal of galaxies close to the origin, the expected
number of galaxy pairs in the absence of clustering $RR$ can be
calculated analytically. We can therefore use the natural estimator $1
+ \xi=DD/RR$ \citep{landy93}, where $DD$ is the measured number of
galaxy-galaxy pairs.

\section{Results}
\label{sec:results}

We have performed 100 runs, each based on a sample of $10^6$ galaxies
drawn from the $z=0$ output of the HV $\Lambda$CDM simulation. For
each unique pair of galaxies within the sample we have selected two
locations for the origin, one where the galaxies subject a separation
angle of $\theta=0.355$, chosen to match figure~1 of \citet{papai08},
and one origin where $\theta=0.71$, twice this angular
separation, to emphasise the wide angle effects on pairs separated by a very wide angle. 

We include all pairs with redshift-space separation $x_s<200\mpcoh$,
and each pair was weighted as described in Section~\ref{sec:rsdist},
based on its real-space separation and angle to the line of sight.
Given that, as one can see from Eq.~(\ref{eq:xi1}), the mode-coupling
terms depend strongly on the radial galaxy distribution, for both the
aperture angles we selected samples with different radial galaxy
distributions, in order to test our methodology in different cases; we
chose $\alpha=0,0.5,2,4$:
\begin{itemize}
\item $\alpha=0$ corresponds to the case of a galaxy distribution that
  has equal density in radial bins of equal width. In this case there are no mode-coupling
  terms  in Eq.~(\ref{eq:xi1}), so this corresponds to the "pure wide angles" case.
\item $\alpha=0.5$ corresponds to a final real-space distribution of
  galaxy pair separations with probability density function
  proportional to $x^2dx$. This is the same distribution obtained by
  randomly sampling pairs from the simulation, so the $x$ factor drops
  out of Eq.~(\ref{eq:weight}). Note that, for clarity, we do not
  present results from this value of $\alpha$ as they are very similar
  to, and overlap results for $\alpha=0$. The match between data and
  theory has the same quality as for the other values of $\alpha$.
\item $\alpha=2$ matches a distribution of galaxies that are uniformly
  distributed in volume, so equal volumes contain equal numbers of
  galaxies. Most planned surveys aim to observe galaxies with this
  radial distribution. Note that, for this galaxy distribution, the
  pair distribution goes as $x^5 dx$.
\item $\alpha=4$ represents a steeper radial distribution where more
  galaxies are found at larger distances. This will increase the
  effect of the mode-coupling terms in the wide-angle redshift-space
  distortion formulae (Eq.~\ref{eq:weight}).
\end{itemize}

\begin{figure}
\centering
\includegraphics[width=1 \columnwidth]{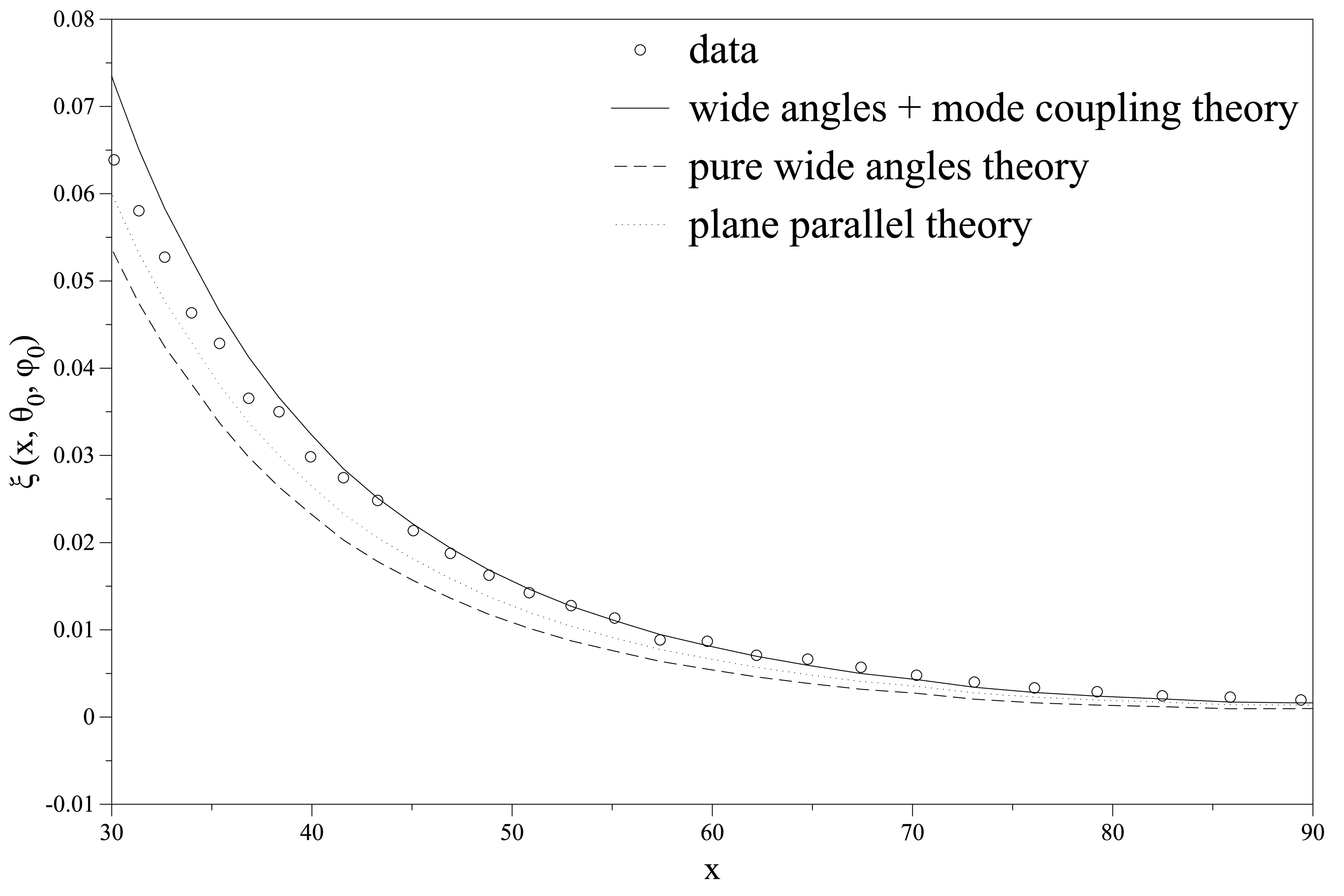}
\caption{Correlation function as a function of scale for fixed
  $\theta=0.355$ and for a bin centered at $\mu = 1.03$, designed to
  match figure~1 of \citet{papai08}. The plane-parallel model is shown
  by the dotted line, the pure wide-angles model by the dashed line
  and the full mode-coupling prediction with $\alpha=2$ by the solid
  line. We expect the match between data and models to be good at
  scales larger than $40\mpcoh$, because we didn't model the small scale
  non linearities. \label{fig:xi_cmpr_papai} }
\end{figure}

We have calculated errors by comparing outputs from 100 subsamples,
that cover the same HV volume and so only contain the shot noise
element.  These errors will therefore underestimate the true error,
because they do not fully include the cosmic variance
component. However, the HV volume is $(3000\mpcoh)^3$, and we only
consider pairs with $x<200\mpcoh$, so we expect the shot noise to
dominate the error budget. Even so, it is worth pointing out that our
primary aim is to consider deviations from plane-parallel theory and
that our proposed methodology works to match data with the full
mode-coupling theory: the size of the errors is unimportant, provided
they are far smaller than the differences between theories.

Fig.~\ref{fig:xi_cmpr_papai} shows the correlation function,
calculated within a narrow bin in $\mu$ for $\theta=0.355$ and
$\alpha=2$; this is equivalent to figure~1 of \citet{papai08}. We are
able to fit the theoretical correlation function to the estimate from
HV simulations for scales larger then $40\mpcoh$ (we do not expect to
match perfectly data with theory on smaller scales because we didn't
model non-linearities).  Looking at plane-parallel, pure wide-angle
and full mode-coupling theories, it is clear that only the full
mode-coupling theory provides a good fit to the data.

\begin{figure}
\centering
\includegraphics[width=1 \columnwidth]{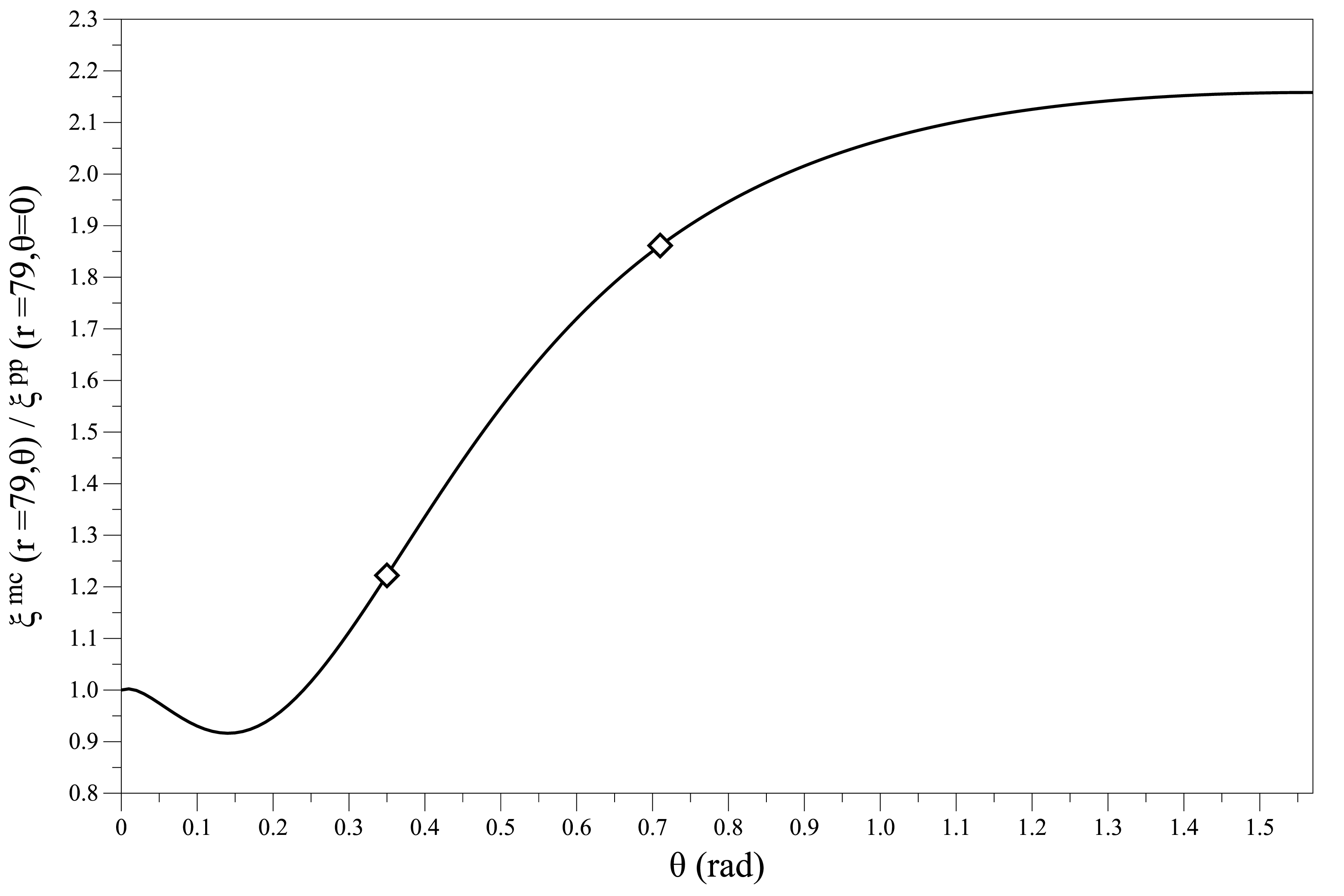}
\caption{Ratio of theoretical full mode-coupling correlation function
  to plane-parallel correlation function, computed for $\alpha=2$ at
  fixed value of $r=79\mpcoh$; the points mark the values of
  $\theta$ used in our simulations. \label{fig:ratio} }
\end{figure}
Fig.~\ref{fig:ratio} shows the ratio between the correlation function
predicted by the full mode-coupling theory and that for the
plane-parallel case, as a function of $\theta$, for a fixed value of
$r=79\mpcoh$, with $\alpha=2$; from this plot we can see that even
for small angles there is a non negligible difference.

In order to analyse RSD, the measured correlation function is usually
decomposed into Legendre momenta, which contain all of the RSD signal
\citep{hamilton98}. Some combination of these momenta is then used to
constrain cosmological parameters through their effects on
the growth of structure. To see how wide angle effects would modify
these measurements we estimate first three even Legendre momenta of
the correlation function from HV mock catalog. We measure:
\begin{equation}  \label{eq:xitildadata}
  \tilde{\xi}_{\rm \ell}(x) = \frac{\displaystyle\sum_\mu DD(x,\mu) 
    L_{\rm \ell}(\mu)}{\displaystyle\sum_\mu RR(x,\mu)},
\end{equation}
for $\ell = 0, 2, 4$. Adopting the plane-parallel philosophy
\citep{hamilton92}, one might be tempted to interpret these functions as:
\begin{eqnarray}
  \xi_{0}^{\rm pp}(x) &=& 
    \left(1 + \frac{2}{3}\beta + \frac{1}{5}\beta^2\right)\xi(x),\\
  \xi_{2}^{\rm pp}(x) &=& 
    -\left(\frac{4}{3}\beta + \frac{4}{7}\beta^2\right)\xi(x),\\
  \xi_{4}^{\rm pp}(x) &=& \frac{8}{35}\beta^2\xi(x),
\end{eqnarray}
where $\xi(x)$ is a real space correlation function and $\beta =
f/b$. In fact the functions in Eq.~(\ref{eq:xitildadata}) are given by:
\begin{equation}
  \tilde{\xi}_{\rm \ell}(x) 
  = \displaystyle\int W_{\rm r}(\mu)\xi^s(x,\theta,\mu)L_{\rm \ell}(\mu)d\mu,
\end{equation}
\noindent
where $W_{\rm r}(\mu)$ is a weight function that appears because of
geometrical constraints and because the distribution of $\mu$ will not
be uniform. We do not correct for this weight in our measurement from
simulations, and apply it to the theoretical calculations that we
compare against.

\begin{figure*}
\centering
\includegraphics[width=1\columnwidth]{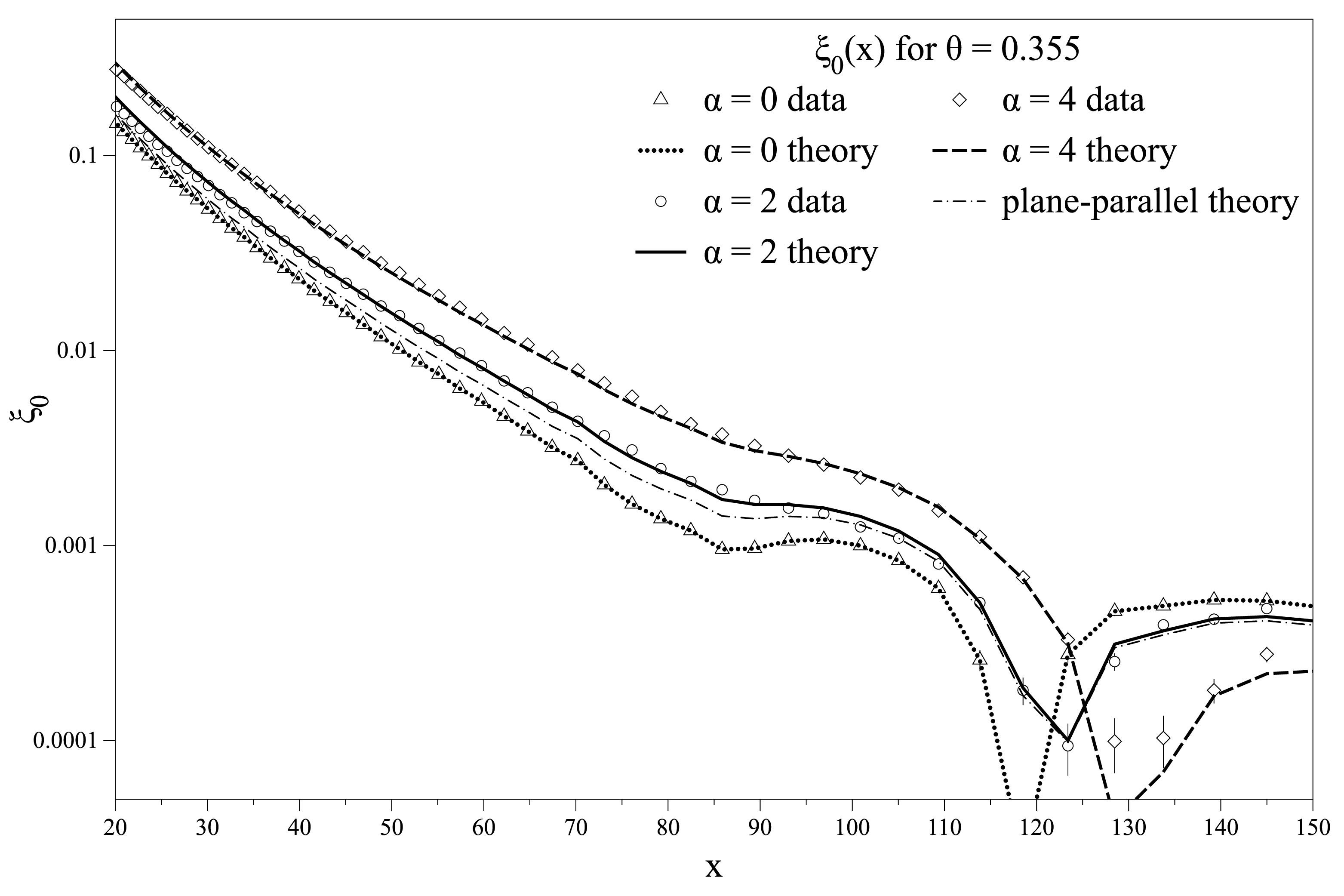}
\hfill
\includegraphics[width=1\columnwidth]{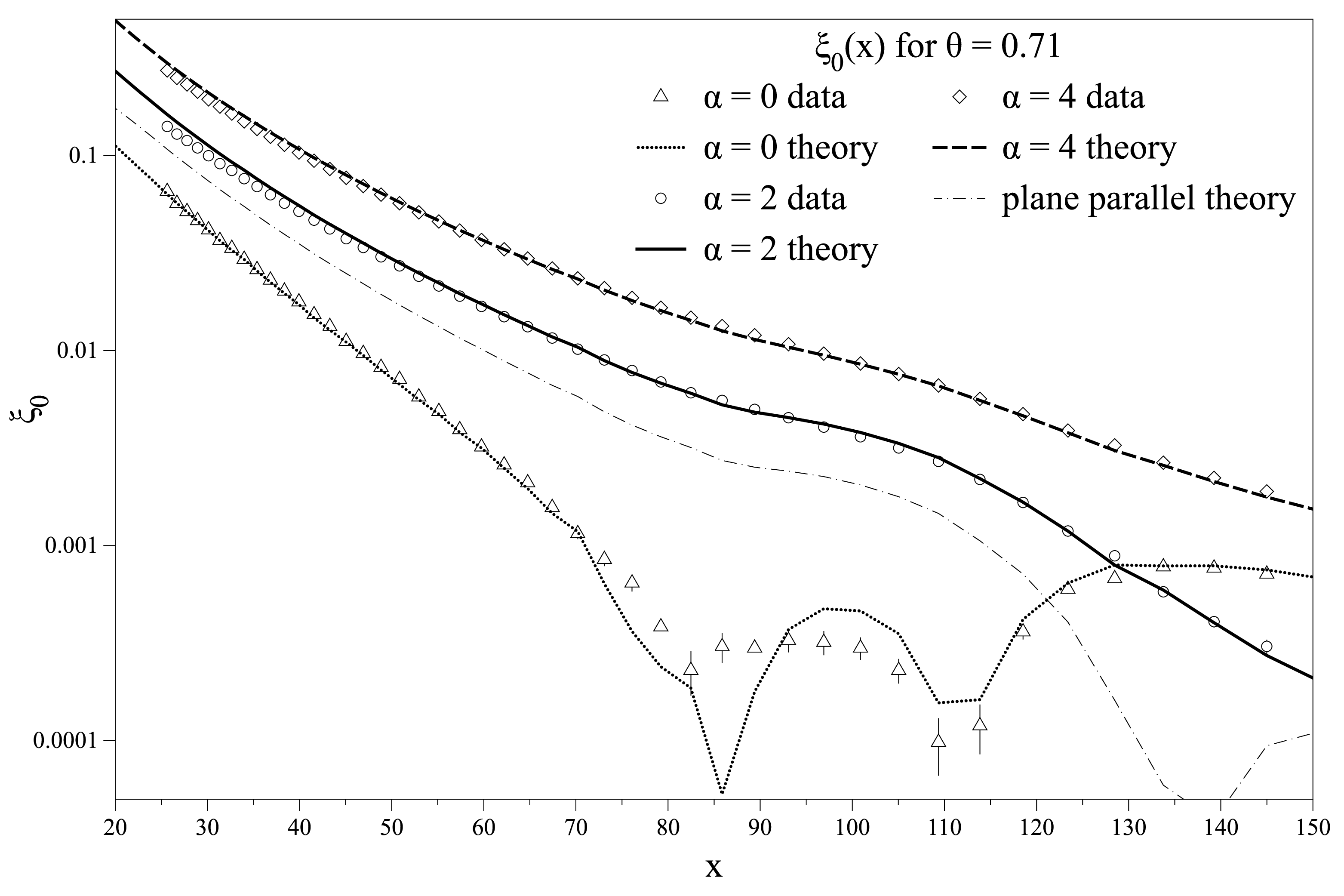} \\
\includegraphics[width=1\columnwidth]{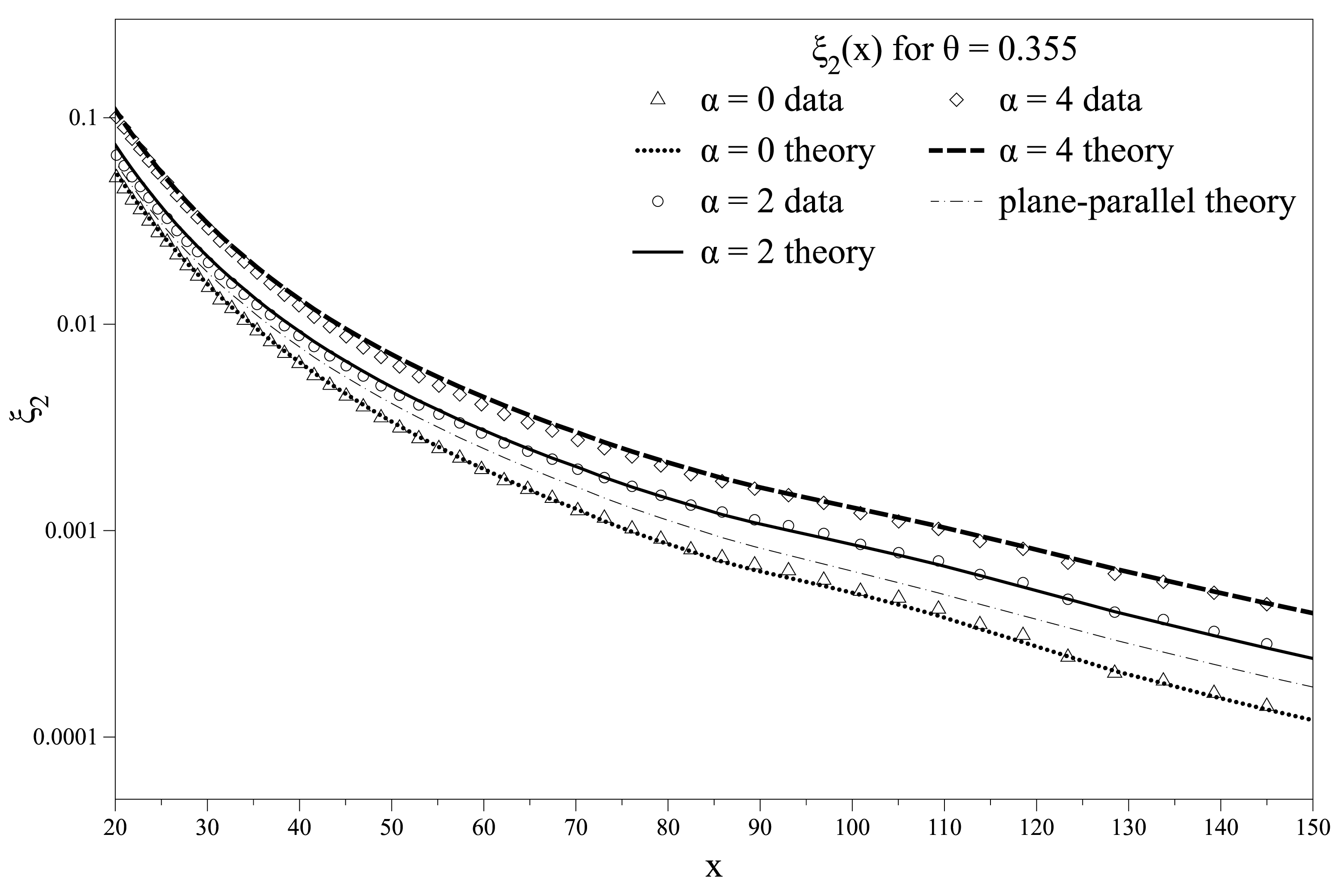}
\hfill
\includegraphics[width=1\columnwidth]{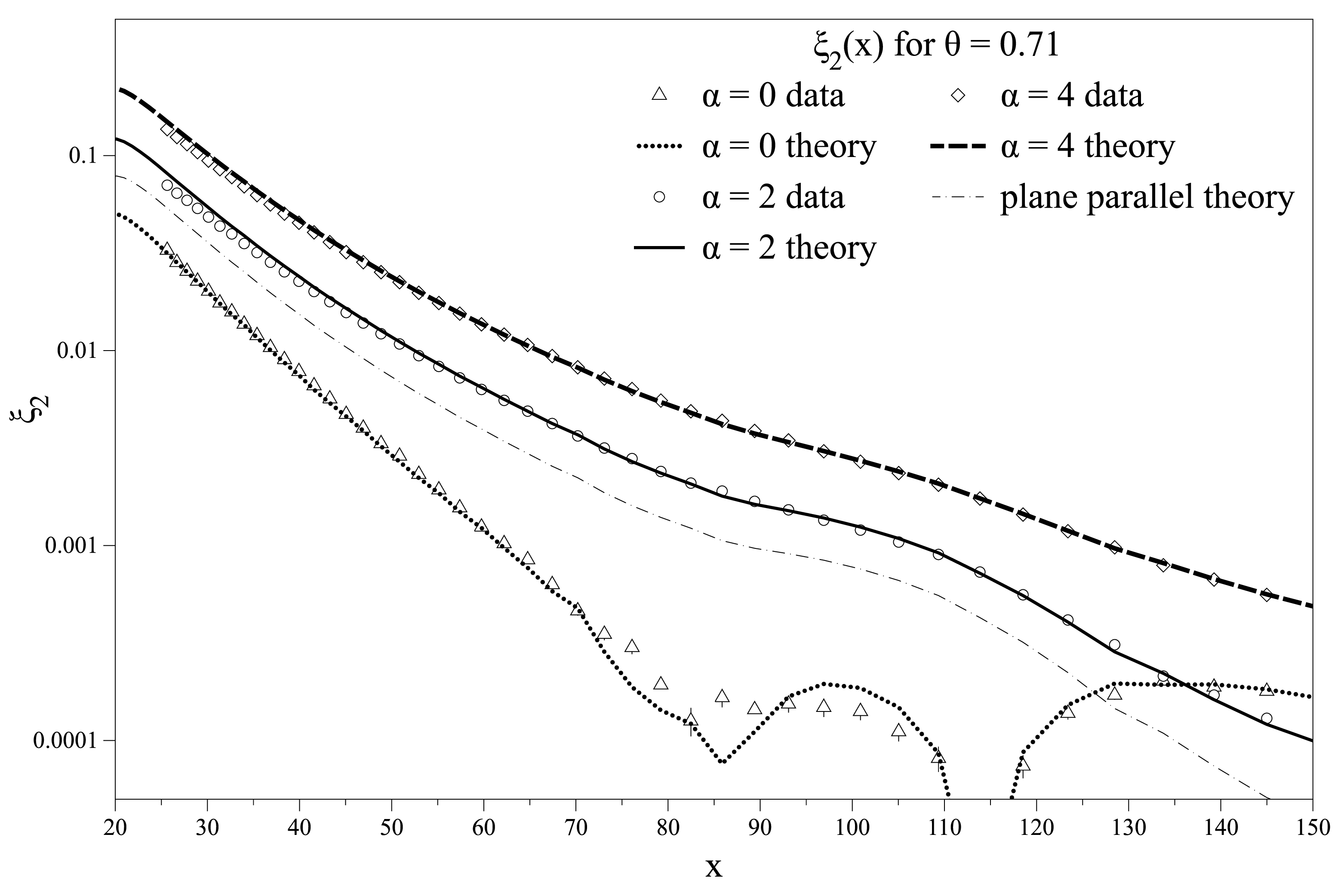} \\
\includegraphics[width=1\columnwidth]{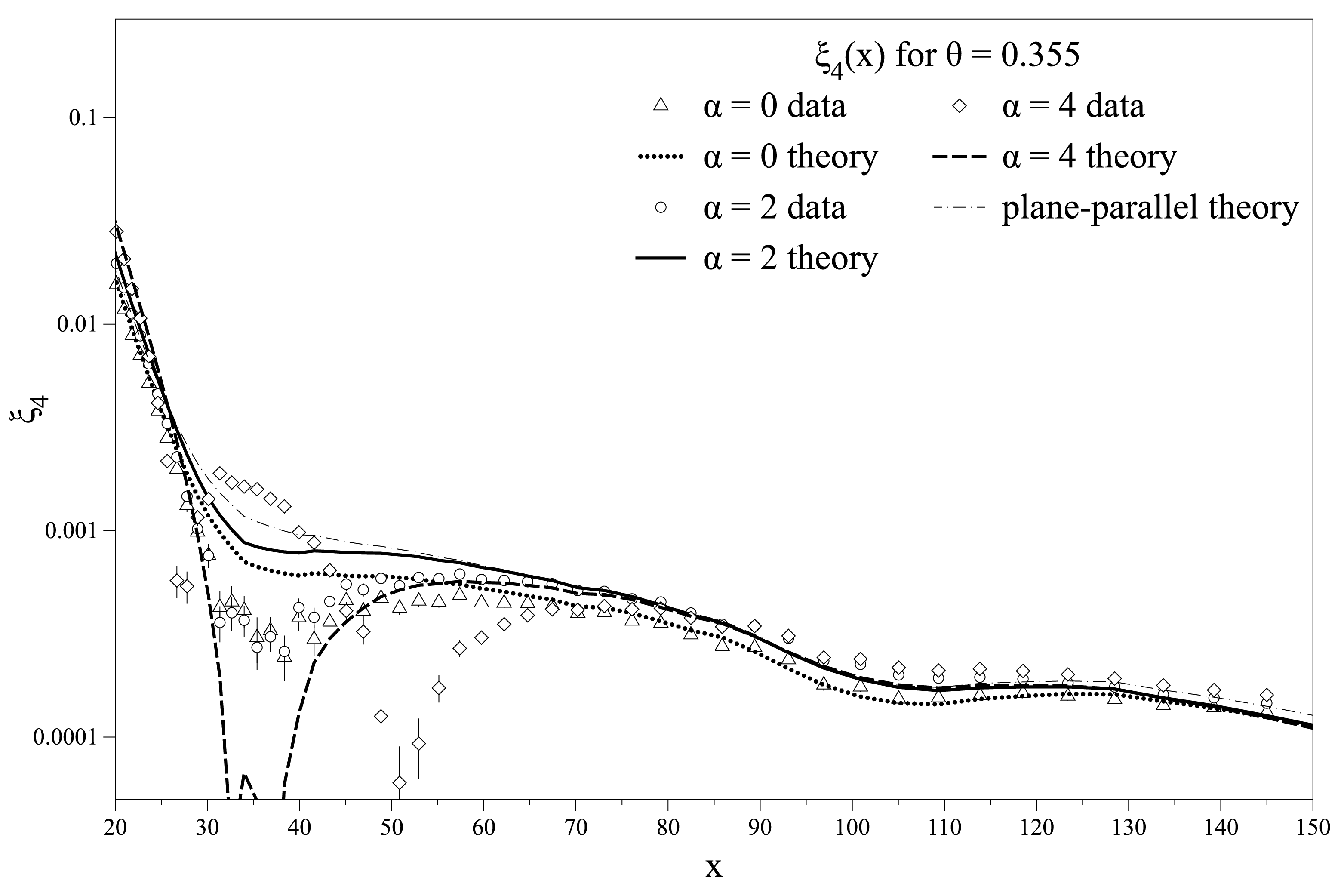}
\hfill
\includegraphics[width=1\columnwidth]{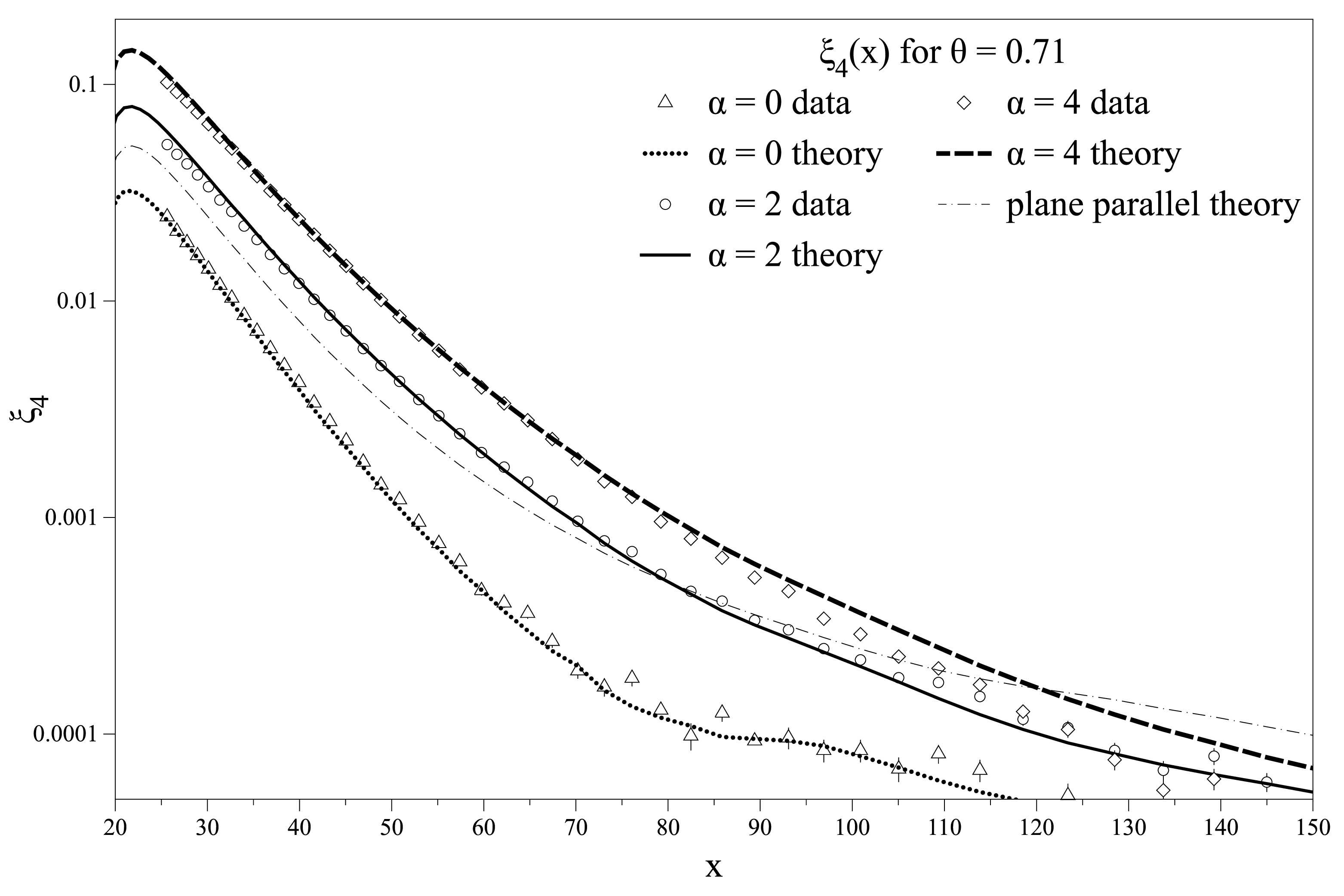} \\
\caption{$\xi_0(x)$, $\xi_2(x)$, $\xi_4(x)$ for angular galaxy pair
  separations of $\theta=0.355$ (left column) and $\theta=0.71$ (right
  column) for different power law radial galaxy density distributions; lines are
  the expected values from the theory of \citet{papai08} with wide
  angle and mode-coupling effects, with $\alpha=0$ (dotted), $\alpha=2$ (solid), 
  and $\alpha=4$ (dashed); dot-dashed lines are plane parallel predictions. 
  Symbols were measured from the HV mocks, with $\alpha=0$ (triangles), 
  $\alpha=2$ (circles), and $\alpha=4$ (diamonds). \label{fig:xil_cmpr} }
\end{figure*}
Fig.~\ref{fig:xil_cmpr} presents the main results of this work,
showing the application of our methodology to sample and analyse data,
and comparing against the full mode-coupling predictions.  We plot
$\xi_0(x)$, $\xi_2(x)$, $\xi_4(x)$ for both the aperture angles and
for radial galaxy density distributions corresponding to $\alpha=0$,
$\alpha=2$ and $\alpha=4$; as we can see, the contribution of the mode
coupling terms increase with $\alpha$, and as the radial galaxy
distribution steepens, more galaxies are moved nearer by the RSD,
leading to an increase in the small-scale correlation
function. Momenta of the correlation function computed with the full
mode coupling theory match remarkably well the data analysed as
explained in Sec.~\ref{sec:rsdmeas}. We also plot the plane-parallel
prediction as a comparison, and this demonstrate how badly this
approximation fails for galaxy pairs with wide angular separation. The
only parts where our methodology do not match the mode coupling theory
are where the correlation multipoles are very small (around $10^{-4}$)
or for small scales, where non-linearities become important; we also
have to take in account that, for the steeper galaxy distributions, we
are upweighting results from the same number of pairs as the more
shallow galaxy distributions so, although the signal is stronger, the
relative error will be the same.

\section{Discussion and conclusions}
\label{sec:disc}

Redshift-Space Distortion Analyses are a powerful tool for cosmology,
but incoming data need to be analysed very carefully in order to fully
extract all available information. Until now RSD analyses have
concentrated on using the plane-parallel approximation (with a few
exceptions including \citealt{okumura08,pope04,matsubara04}); this is
almost correct if the survey is narrow, but when galaxy surveys with
wide field of view data will be available, there will be a consistent
number of galaxy pairs separated by wide angles and in this case the
plane-parallel approximation fails.

This approximation arises from not taking in account some of the terms
in the Jacobian relating redshift- to real-space; in this paper we
considered its exact expression, and this causes additional,
non-diagonal terms in the correlation function. We then use the
expansion of the correlation function in a base of tripolar spherical
harmonics, as suggested in \citealt{papai08}; following the formalism
developed in \citealt{szapudi04, papai08}, we make predictions for the
momenta of the correlation function for galaxy pairs at fixed angular
separation, and we test them against data from the Hubble Volume
Simulation.

In order to do this we have had to introduce a
new methodology: rather than creating a single sample of galaxies in
redshift-space from which we can count pairs, we have instead
dynamically applied RSD on a pair-by-pair basis, choosing an origin
for each. By including weighting functions in $x$ and $\mu$ we can
match results from the more traditional approach. This allows us to
only consider galaxy pairs at particular values of $\theta$.

To show both the correctness of our methodology and the deviation from
the plane-parallel situation we tested galaxies separated by two
different fixed values of $\theta$; the mode coupling terms are also
strongly dependent on the galaxy radial distribution, so we have
tested simulations with 4 different number density distributions, for
both values of $\theta$.

We show that taking in account the wide angle and mode coupling terms
(that are of the same order, as stated in \citealt{papai08}) give a
clear deviation from the plane-parallel theory; using the exact theory
and our methodology, we can match the results of simulations, the
agreement between data and theory being remarkable, especially
considering how crude our modelling of the HV simulations is (we use a
measurement of the 3D real-space correlation function as our baseline
model, and we do not include a correction for fingers-of-god type
effects).  For RSD measurements made within radial bins, it will be
vital to match the theory to the exact distribution of galaxies
observed.

In a measurement of RSD from real data, the final result will be a
weighted average over different opening angles, to take account for
the fraction of galaxy pairs separated by a $\theta$ angle, so in wide
surveys one has to discard the plane-parallel approximation, or face
losing a considerable amount of information. For this reason our
methodology will be particularly useful for incoming and future
redshift Surveys, such as the Baryon Oscillation Spectroscopic Survey
(BOSS; \citealt{schlegel09a}), BigBOSS \citep{schlegel09b}, and Euclid
\citep{laureijs09}; the importance of this methodology for current and
future experiments will be considered in a subsequent paper
\citep{samushiaprep}.

\section{Acknowledgements}

AR is grateful for the support from a UK Science and Technology
Facilities Research Council (STFC) PhD studentship. WJP is
grateful for support from the European Research Council and from the Leverhulme Trust and STFC. 
LS acknowledges support from European Research Council, 
Georgian National Science Foundation grant ST08/4-442 and SNSF (SCOPES grant No 128040).
AR would like to thank G. W. Pettinari for the useful discussions. Simulated
data was calculated and analysed using the COSMOS Altix 3700
supercomputer, a UK-CCC facility supported by HEFCE and STFC in
cooperation with CGI/Intel.

\label{lastpage}

\end{document}